\documentclass[aps,prl,twocolumn,floatfix,nofootinbib]{revtex4}

\usepackage{psfrag,graphicx}
\usepackage{dcolumn}
\usepackage{amsmath,amssymb}
\usepackage{bm}
\usepackage{amsthm}
\theoremstyle{plain}
\newcommand{\Tr}{{\text Tr}}
\newcommand{\half}{\mbox{$\textstyle \frac{1}{2}$}}
\newcommand{\ket}[1]{\left | #1\right \rangle}
\newcommand{\bra}[1]{\left \langle #1\right |}

\begin{document}

\title{Optimal purification of thermal graph states}

\author{Alastair Kay and Jiannis K. Pachos}
\affiliation{Department of Applied Mathematics and Theoretical
Physics, University of Cambridge, Cambridge CB3 0WA, UK}
\author{Wolfgang D\"ur and Hans-J. Briegel}
\affiliation{Institut f\"ur Quantenoptik und
Quanteninformation der \"Osterreichischen Akademie der
Wissenschaften, Innsbruck, Austria\\ Institut f\"ur Theoretische
Physik, Universit\"at Innsbruck, Technikerstra\ss e 25, A-6020
Innsbruck, Austria}

\date{\today}

\begin{abstract}

In this paper, a purification protocol is presented and its performance is
proven to be optimal when applied to a particular subset of graph states
that are subject to local $Z$-noise. Such mixed states can be produced by
bringing a system into thermal equilibrium, when it is described by a
Hamiltonian which has a particular graph state as its unique ground state.
From this protocol, we derive the exact value of the critical temperature
$T_\text{crit}$ above which purification is impossible, as well as the
related optimal purification rates. A possible simulation of graph
Hamiltonians is proposed, which requires only bipartite interactions and
local magnetic fields, enabling the tuning of the system temperature.

\end{abstract}


\maketitle

{\em Introduction.} Any quantum technological implementation is plagued by
environmental noise. The possibility to purify quantum states, or to use
error correcting algorithms to stabilize quantum operations, is therefore a
necessary step towards reaping the benefits of quantum technologies. Much
attention has been focused lately on the
purification~\cite{murao:98,Ma00,Lo04,Duer,aschauer:04,Goyal,Briegel:06} of
a large class of multipartite states called graph states~\cite{He06}. After
the initial restriction to two-colourable graph states
\cite{Duer,aschauer:04}, the ideas have been extended to all graph states
\cite{Briegel:06}, and generalized to other stabilizer states
\cite{Knill:06}. The variety of different protocols trade off between a
large tolerance to noise \cite{Duer,aschauer:04} and the rate of
purification \cite{Goyal}. Graph states have proven important for realizing
a variety of quantum information tasks such as performing quantum
computation \cite{Raussendorf}, quantum communication
\cite{christandl-2005-3788} and as a means for efficiently approximating
other quantum states, such as the ground states of strongly correlated
systems \cite{Anders}. While some bounds
have previously been proven on the ability to purify multipartite
states \cite{dur:99,purify_thermal}, optimal results only exist for two-qubit
states \cite{bipartite_purification:2,bipartite_purification:3}.

Here we concentrate on the purification of graph states that are
subject to the physically motivated independent
$Z$-noise. From the technological perspective, large-scale quantum
computation is still too difficult to implement. However, graph states
can be made and manipulated in the laboratory,
e.g.~by controlled collisions of alkali atoms trapped in optical lattices~\cite{Mandel,Alastair1}. To enable the controlled collisions, it is necessary to employ two magnetically sensitive hyperfine levels of the atoms to form a qubit.
This sensitivity means that the states are subject to decoherence from stray magnetic fields, in addition to any uncontrolled collisions that may occur. These errors are described by $Z$-errors, and as experiments improve, one can expect them to become more localised.
 An alternative approach to preparing a graph state involves implementing a Hamiltonian, known as a graph Hamiltonian, which has the desired state as its ground state. As it is
impossible to cool the system to absolute zero, the resulting equilibrium state will always be a thermal state. For graph
Hamiltonians, the thermal noise corresponds to local $Z$-noise on
each qubit.

In this paper, we consider a certain purification protocol applied to
arbitrary graph states in the presence of independent $Z$-errors. Whilst
this multipartite protocol is not novel or sophisticated
\cite{murao:98,dur:04,purify_thermal}, it has the advantage of being analytically
tractable. Most importantly, we prove optimality of this protocol for these
types of error, both in the sense of the level of noise that can be
tolerated as well as the scaling of the purification rate, for a specific
subset of states that includes the cluster and Greenberger Horne Zeilinger
(GHZ) states. For up to seven qubits, this subset can be shown to be
isomorphic to arbitrary graphs under local operations. We propose a method
for simulating the graph Hamiltonians that is comprised of only two-body
collisions and local magnetic fields, even though the resulting Hamiltonian
has at least 3-qubit interactions \cite{Browne,nielsen:05}. The critical
purification temperature related to these models can easily be made to lie
above the typical temperatures given, e.g. from optical lattice realizations
of such Hamiltonians.

{\em Graph Hamiltonians and Graph States.} Let us introduce a graph $G$,
which is defined by a set of vertices $V_G$, and a set of edges $E_G$,
describing the connections between the vertices. To each vertex of this
graph, we attach a spin-\half\;particle (qubit), and define a graph state to be the
ground-state of the following Hamiltonian,
\begin{equation}
H=-\half\sum_{i\in V_G}B_iK_i ,\,\,\,\,\,K_i=X_i\!\!\prod_{\{i,j\}\in
E_G}\!\!Z_j.
\label{Hamiltonian}
\end{equation}
The $B_i$'s are the coupling strengths, which we henceforth take to have
equal magnitude, $B_i=B$, and we assume $B>0$. The interaction terms $K_i$
commute with each other, $[K_i,K_j]=0$, and hence each term individually
stabilizes the eigenstates of $H$. A local Pauli $Z$-rotation on qubit $i$,
$Z_i$, commutes with all the $K_j$ where $j\neq i$, and anti-commutes with
$K_i$. Hence, the excitations of the Hamiltonian are given by local
$Z$-rotations applied to the ground state.

A constructive way to produce graph states is found by close analogy with
the cluster states~\cite{Browne,He06}. We can produce a general graph state
by creating the $\ket{+}=(\ket{0}+\ket{1})/\sqrt{2}$ state on each vertex of
$G$, and applying controlled-phase gates along the edges of the graph. The
action of measurements on the cluster state also applies to the graph
states, such that when a $Z$-measurement is performed, all nearest-neighbour
bonds are severed. This means that graph states can be cut into sections
using $Z$-measurements, and can be combined together using controlled-phase
gates. The key to what follows is the realization that $Z$-errors commute
with both of these operations, and consequently remain as $Z$-errors.

Examples of graph states that are of particular interest in quantum
computation and communication are the cluster~\cite{Br01} and GHZ~\cite{GHZ}
states. The first state corresponds to a graph that is given by a
$d$-dimensional cube, while the latter state corresponds to a graph that is
locally equivalent to a single qubit connected with all other qubits.
\begin{figure}[!t]
\begin{center}
\includegraphics[width=0.4\textwidth]{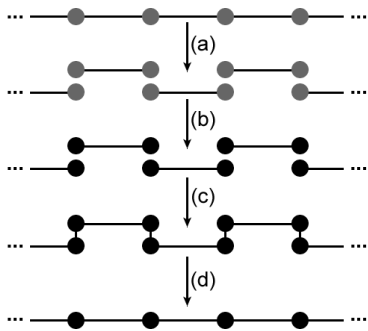}
\end{center}
\caption{Purification protocol for independent noise. First, we take
  many copies of the noisy graph state. (a) We construct from these
  two-qubit nearest-neighbour states (noisy). (b) Two-qubit states are
  purified (if possible). (c) Controlled-phase gates are applied between
  local qubits. (d) All qubits except one from each party are measured in
  the $X$-basis, leaving the remaining qubits in the purified state.}
\vspace{-0.5cm}
\label{fig:protocol}
\end{figure}

{\em Purification of Thermal Graph States.} The thermal state of the
Hamiltonian in Eqn.~(\ref{Hamiltonian}) of $|V_G|=N$ qubits at temperature
$T=1/(k_B\beta)$ is given by
\begin{equation}
\label{thermalstate}
\rho=\frac{e^{-\beta H}}{\Tr(e^{-\beta H})},
\end{equation}
where we set the Boltzmann constant $k_B$ equal to unity. This density
matrix can be written in terms of local $Z$-errors as $\rho(p) = {\cal E}_1
{\cal E}_2 ...{\cal E}_N
\ket{\psi}\bra{\psi}$, where
\begin{equation}
{\cal E}_i \rho =\left[(1-p)\rho + pZ_i\rho Z_i\right]
\nonumber
\end{equation}
and
\begin{equation}
p=\frac{1}{1+e^{\beta B}}
\label{propability}
\end{equation}
is the probability of a $Z$-error occurring at a certain site
due to the non-zero temperature $T$. The graph state $\ket{\psi}$ is the unique
ground state of Hamiltonian
(\ref{Hamiltonian}). Our aim is to purify towards the state
$\ket{\psi}$ using many
copies of $\rho(p)$. We consider that each vertex of the graph is controlled
by a different party and that operations such as measurements and
controlled-phase gates are only allowed locally, but involving the many copies. This restriction corresponds to the scenario of quantum repeaters \cite{bipartite_purification:1}, where the different parties are physically separated, and also serves to illustrate the entanglement properties of the system. We do not envisage implementation of the protocol in other scenarios, since it is generally cheaper in terms of resources to create the state directly -- our aim is to prove optimality, providing a benchmark for all other protocols.

The purification procedure consists of breaking down the graph states into
smaller blocks, purifying them, and then recombining them. We refer to it
as the Divide and Rebuild Purification Protocol (DRPP). The smaller blocks
that we choose to use are two-qubit states, for which there are analytic
purification results, returning maximally entangled states. As already
specified, the splitting of the graph state is readily achieved with
$Z$-measurements. This leaves us with a two-qubit mixed state $\rho_2(p)$,
which can be purified to a maximally entangled state $\ket{\psi_2}$ provided
\begin{equation}
\bra{\psi_2}\rho_2(p)\ket{\psi_2}>\half.
\nonumber
\end{equation}
Once we have generated a maximally entangled state for each of the edges of
the graph, each local party ($i$) holds a number of qubits equal to the
number of nearest-neighbours $|E_G^i|$. These can be reduced to a single
qubit that contains all these links by applying
controlled-phase gates and performing $X$-measurements, as described
in Fig.~\ref{fig:protocol}.

The
condition for purification of the two-qubit state between neighbouring
sites is readily found to be
\begin{equation}
(1-p)^2>\half. \label{eqn:condition}
\end{equation}
This is known to be necessary and sufficient for purifying states that are
Bell diagonal, like the states considered here
\cite{bipartite_purification:2,bipartite_purification:3,bipartite_purification:1}. We
can hence calculate that the maximum temperature at which the DRPP can
purify graph states is given by
\begin{equation}
T_\text{crit}=\frac{-B}{\ln(\sqrt{2}-1)}.
\label{critical}
\end{equation}
\begin{figure}[!t]
\begin{center}
\includegraphics[width=0.4\textwidth]{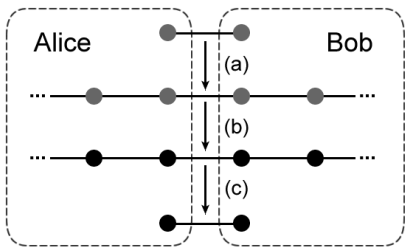}
\end{center}
\caption{If we assume the multipartite state can be purified,
then this implies that we can purify the two-qubit state. Conversely,
if the two-qubit state cannot be purified, the assumption must be
broken. (a) Alice and
Bob take the two-qubit state and reconstruct the noisy graph state. (b) This
state is purified by assumption. (c) All extra qubits are
measured out to return the original pair, now pure.}
\vspace{-0.5cm}
\label{fig:optimality}
\end{figure}

{\em Optimality and Rates.} While the DRPP can be applied to any graph state
(with minor modifications), we can prove its optimality for a specific
subset of graphs. For clarity, we shall restrict to only cluster states (of
arbitrary dimension) and GHZ states, all of which fall into this
classification. The optimality that we prove is with respect to both the
level of noise that can be tolerated such that purification is still
possible, and with respect to the number of copies of the initial, noisy,
state required to form a single pure copy. In particular, one can prove that
states with higher levels of noise than the one dictated by
Eqn.~(\ref{eqn:condition}) can never be purified with {\em{any}} protocol.
We do this by considering a purification protocol for a two-qubit state. The
two parties sharing this state are allowed to introduce extra qubits and the
local operations that they apply subsume multipartite considerations for the
additional qubits.

\begin{figure}[!t]
\begin{center}
\includegraphics[width=0.4\textwidth]{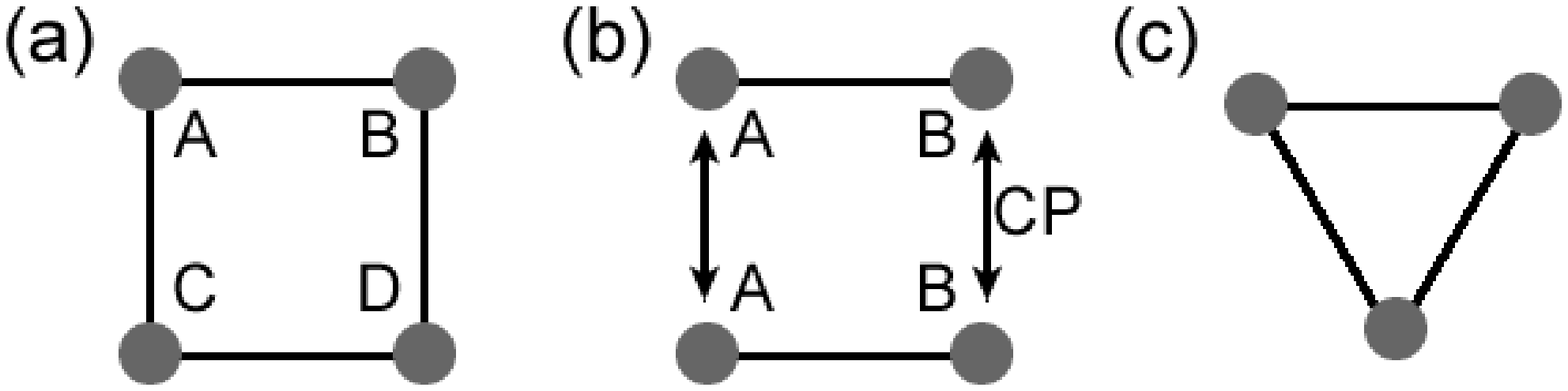}
\end{center}
\caption{(a) Simplest example of a $2d$-cluster state, shared between 4 parties. (b) Alice and Bob
can locally reconstruct the square graph using two copies of $\rho_2$, and
applying controlled-phase gates between them. (c) For the triangular
configuration, the optimality proof fails.}
\vspace{-0.5cm}
\label{fig:graphs}
\end{figure}

If two parties, Alice and Bob, hold several copies of the noisy two-qubit
state $\rho_2(p)$~\cite{Comment}, they can locally recreate the initial
thermal state, $\rho(p)$. For linear graphs (Fig.~\ref{fig:optimality}),
this simply corresponds to Alice and Bob locally creating their own thermal
cluster states, and connecting them to $\rho_2(p)$ with controlled-phase
gates. In the case of a more general graph, we require some additional
connections between Alice and Bob. These are achieved by using multiple
copies of $\rho_2(p)$ (Fig.~\ref{fig:graphs}(b)). At this stage, we assume
that purification of $\rho(p)$ is possible, yielding $\ket{\psi}$. From
there, Alice and Bob can measure out all the qubits that they added, leaving
a pure two-qubit state. Hence, if $\rho(p)$ can be purified, $\rho_2(p)$ can
always be purified. However, we know that $\rho_2(p)$ cannot be purified if
$(1-p)^2\leq\half$. Hence, under this condition, our assumption must be
false i.e.~the multipartite state $\rho(p)$ cannot be purified, whatever the
protocol. The DRPP saturates this bound and hence is optimal for independent
$Z$-noise. While analysis of the protocol of \cite{Duer,aschauer:04} is a difficult problem, numerical results indicate that it also saturates the bound, and hence is optimal for this type of noise.

The rate of purification, $R_\psi$, of the graph state $\ket{\psi}$ can
be calculated in terms of the yield of a Bell state, $R_2$. We take the
standard definition of rate,
$$
R_\psi=\frac{\rm{Copies\; of\; } \ket{\psi} \rm{\;
produced}}{\rm{Copies\; of\; } \rho \rm{\; consumed}}
$$
If we can purify $\rho$ into $\ket{\psi}\!\!\bra{\psi}$ at a rate $R_\psi$
(i.e.~we require $1/R_\psi$ copies of $\rho$ to create $\ket{\psi}$), then
we can create a Bell state between any linked pair just by performing
$Z$-measurements on excess qubits. There could be a more efficient way to
generate this Bell pair, requiring fewer copies, so $R_\psi\leq R_2. $

Similarly, if we can purify Bell pairs, then we can generate $\ket{\psi}$.
For that we take a Bell pair between each nearest-neighbour and the local parties
perform the reconstruction as specified by the DRPP
(e.g.~controlled-phase between local states and
$X$-measurements). Most of the Bell pairs can be purified in parallel --
we only require $N_\text{geo}$ copies of $\rho$ to purify enough
copies, where the geometric factor $N_\text{geo}$ depends only on the local degree of the graph, $|E_G^i|$, and is
otherwise independent of the number of qubits in the graph.
This is because the $Z$-measurements commute with the errors, and divide the
state $\rho$ into separate blocks. As a result, $ R_\psi\geq
\frac{R_2}{N_\text{geo}}. $ Combining the two results,
$$
R_2\geq R_\psi\geq \frac{R_2}{N_\text{geo}}
$$
\begin{figure}[!t]
\vspace{0.5cm}
\begin{center}
\includegraphics[width=0.4\textwidth]{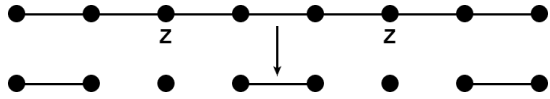}
\end{center}
\caption{By measuring in the $Z$ basis on every third qubit, we can create
two-qubit states between all nearest-neighbours with only three copies of the
original linear cluster state. Hence, we say $N_{\rm{geo}}=3$.}
\vspace{-0.5cm}
\label{fig:geo}
\end{figure}

For $d$-dimensional cluster states, we can readily evaluate $N_\text{geo}$,
since we must measure out all qubits connected directly to the pair that we
are interested in isolating. Starting from the edge of the lattice, this
uses up $3d-2$ qubits (we include in this number some single qubits that,
while we might not intend to measure them, become isolated), plus the
two for the state $\rho_2$. If our lattice extends to $N$ qubits in each
direction, then we can generate on average $(N-1)N^{d-1}/(3d)$ copies of
$\rho_2$ from a single copy of $\rho$. We need $d(N-1)N^{d-1}$ different
copies of $\rho_2$, and hence $N_\text{geo}=3d^2$. As demonstrated in
Fig.~\ref{fig:geo}, this corresponds to $N_\text{geo}=3$ for $d=1$. The
resulting rate is independent of the number of qubits in the system, only
depending on $R_2$, which is optimal up to a small numerical factor. For
$N$-qubit GHZ states, $N_\text{geo}=N-1$.

{\em Physical Implementation.} A significant achievement in recent quantum
engineering experiments is the construction of cluster states with optical
lattices \cite{Mandel}. They are produced with a single operational step
independent of the size of the system. It is natural to consider this setup
for studying the purification properties of thermal cluster states. This
requires the implementation of both the cluster Hamiltonian and a
purification protocol in a physical setup. In the following, we will present
a simple way of simulating the Hamiltonian (\ref{Hamiltonian}) for cluster
states, using proven experimental techniques in optical lattices.

When an entangled state decoheres,
there is a characteristic lifetime that determines when the state
becomes separable. On the other hand,
if it is possible to obtain an interaction described by the Hamiltonian that has this entangled
state as a ground state, then provided the energy gap is large enough
in comparison to the decoherence rate, entanglement can survive
indefinitely in the system, e.g.~in the form of purifiable mixed
states, as we have already seen.

Simulating Hamiltonian (\ref{Hamiltonian}) for a general graph is a
relatively straightforward task. The method we adopt here consists of a
unitary operation, $U_G^\dagger$ [$U_G$], applied before [after] the
evolution with respect to a local Hamiltonian. This evolution is generated by
applying a uniform magnetic field in the $x$-direction. When $U_G$
corresponds to controlled phase gates between all pairs of qubits connected
in the graph, then it is easy to show that the resulting effective
Hamiltonian
\begin{equation}
H= U_G \left(B\sum_iX_i \right)U_G^\dagger
\end{equation}
is of the form (\ref{Hamiltonian}) \cite{Du06}. The spectrum of the
Hamiltonian corresponding to the magnetic field $B\sum_iX_i$ is the same as
that of the Hamiltonian $H$, as they are related by an isospectral
transformation. Hence, the generated thermal state directly corresponds to
the one of $H$.

In optical lattices, the unitary $U_G$ can be realized in cubic lattices by controlled collisions between
nearest-neighbours.
This is precisely the operation which is experimentally employed in
\cite{Mandel} for the generation of a cluster state. Meanwhile, the local magnetic fields are implemented by globally applied Raman transitions between the hyperfine states that encodes a qubit. Thus, the realization
of the graph Hamiltonian is readily achieved. The stationary state of this
system is the thermal state of Eq.~(\ref{thermalstate}). Further, this system gives us the ability to vary the temperature. In previous experiments~\cite{Gatzke} the recorded
temperature after performing optical cooling was given by the relation $T
\approx 0.1U_0$, where $U_0$ is light shift potential created by the optical
lattice. From the value of the critical temperature given in
(\ref{critical}), one deduces that by employing moderate local magnetic
fields with amplitude $B\gtrsim 0.1U_0$, one can bring the system into the
purifiable regime.

Once we are able to implement a purification protocol in optical lattices, whether it be the DRPP or any other, this provides us with the perfect test-bed to probe the maximum temperature that still allows purification, and verify the critical temperature given in Eqn.~\ref{critical}. The potential implementation of such schemes has been described extensively in, for example, \cite{aschauer:04}. The main drawback is the requirement of local addressability, which can be circumvented with the help of superlattices \cite{Alastair1,Alastair2,beige:06} by breaking of the translational invariance of the lattice.

{\em Conclusions.} Here, a purification protocol has been proven to be
optimal when applied to a sub-class of graph states, including the cluster and
GHZ states, subject to $Z$-errors. Although we have restricted ourselves to
this particular form of noise in this paper, we emphasize that thermal
states arising from graph Hamiltonians are precisely of this form, rendering
the considered types of states interesting also from an experimental and
practical point of view.

While the optimality proof cannot be applied to certain types of graphs
(e.g.~the icosahedron), and we are aware of examples of noise (e.g.~local or
global white noise) where the proposed protocol is not optimal, several
extensions of our results are possible. Two of the authors have shown
\cite{in_prep} that the optimality proof can be extended to a wider
class of states and to different forms of noise. For instance, it can be shown that all graph states of up to seven qubits can
be brought by local unitary operations to a form where the optimality proof
can be applied \cite{He06}. Thus, all these graphs have a critical temperature given by
Eqn.~(\ref{critical}). In addition, it shown in \cite{in_prep} how the
optimality proof presented in this paper can be extended to other types of
noise, providing an upper-bound to the error probability that can be
purified. Further extensions taking into account non-graph states and noisy local
operations during the purification protocol will also be
examined.

{\em Acknowledgments.} This work was supported by Clare College, Cambridge (AK), the EPSRC, the FWF, the European Union (OLAQUI,SCALA),
the \"OAW through project APART (WD) and the Royal Society (JKP).


\begin{thebibliography}{27}
\bibitem{Comment} Note that all the noisy two-qubit states are
identical.
\end{thebibliography}
\end{document}